\documentclass[onecolumn,amssymb,amsmath,floats,showpacs,superscriptaddress,pre]{revtex4-1}

\usepackage{graphicx}
\usepackage{dcolumn}
\usepackage{bm}

\usepackage{natbib}
\usepackage[
text={7in,9.5in},centering,
]{geometry}

\usepackage{epsfig}
\usepackage{amsmath}
\usepackage{amssymb}
\usepackage{textcomp}
\begin{document}

\title{A Natural Gauge in Quantum Electrodynamics}

\author{Natalia Gorobey}
\affiliation{Peter the Great Saint Petersburg Polytechnic University, Polytekhnicheskaya
29, 195251, St. Petersburg, Russia}

\author{Alexander Lukyanenko}\email{alex.lukyan@mail.ru}
\affiliation{Peter the Great Saint Petersburg Polytechnic University, Polytekhnicheskaya
29, 195251, St. Petersburg, Russia}

\author{A. V. Goltsev}
\affiliation{Ioffe Physical- Technical Institute, Polytekhnicheskaya
26, 195251, St. Petersburg, Russia}

\begin{abstract}

An alternative representation of the kernel of the evolution operator in quantum electrodynamics is obtained in the form of a functional integral, in which the gauge momentum corresponding to the Gaussian constraint is excluded from the dynamics. The natural gauge condition, that arises as a result of this representation, leaves the integration only over gauge invariant canonical variables.
\end{abstract}


\maketitle

\section{Introduction}
The gauge conditions serve to eliminate the arbitrariness that exists in a singular dynamical theory based on the principle of relativity with a certain continuous symmetry group \cite{1}. Here we include, first of all, electrodynamics and the Yang-Mills theory. In electrodynamics, gauge transformations are gradient transformations \cite{2}. In the classical theory, the choice of the gauge condition is due to considerations of convenience and simplicity of the solution in a particular problem. In quantum theory, we are already talking about eliminating the divergence in the Feynman functional integral that arises when integrating over the orbit of the gauge group \cite{3}, and the gauge conditions should effectively solve this problem. However, in the Yang-Mills theory, whose symmetry group is non-Abelian, anomalies arise \cite{4,5} when the gauge conditions do not provide an unambiguous choice of the orbital element. In this case, the result of the calculations should not depend on the choice of gauge conditions, which is ensured by introducing the Faddeev-Popov determinant in the measure of the Feynman integral \cite{3}. This construction of the covariant quantum theory is formalized within the extended BRST symmetry \cite{6,7} and formulated in the Batalin-Fradkin-Vilkovyssky theorem \cite{8,9}. In \cite{10}, an alternative way to eliminate gauge arbitrariness is proposed, which is a condition for the classicality of the parameters of symmetry transformations in quantum theory. In the case of a free Yang-Mills field, it is shown that additional classicality conditions lead to the separation of the dynamics of the transverse degrees of freedom of the field from the constraint equations that restrict the initial data.

In this paper, the separation of dynamics and initial data is considered using the example of an electromagnetic field interacting with a complex scalar matter field. Here  an additional classicality condition serves as a natural gauge condition, since it is the equation of motion of the collective component of physical degrees of freedom (vector potential and scalar field), which changes under gradient transformations. Note that the proposed method of fixing the (natural) gauge is noncovariant, since, in particular, functional integration over the scalar potential of the electromagnetic field is not assumed. The problem of relativistic invariance in this theory requires additional consideration. However, the same applies to the usual procedure for fixing the gauge conditions \cite{3}. The natural gauge conditions can be expected to be free of anomalies in the case of a non-Abelian gauge group as well.

In the next section, the natural gauge condition is introduced as a modification of the original Lagrange function in electrodynamics. In the third section, it is shown that the Feynman functional integral for the propagator of the modified theory is nonsingular without additional restrictions on the measure of integration.

\section{Classical electrodynamics in natural gage}
The original Lagrange function of an electromagnetic field interacting in a minimal way with a charged scalar field can be written in the form:
\begin{equation}
 L=\frac{1}{2}\int{d^3x}\left\{[(\dot{A}_i-\partial_i{A_0})^2-B_i^2]+[(\partial_0+ieA_0)\overline{\phi}(\partial_0-ieA_0)\phi-(\partial_i+ieA_i)\overline{\phi}(\partial_i-ieA_i)\phi-V(\overline{\phi}\phi)]\right\},
 \label{1}
\end{equation}
where $B_i$ is the magnetic induction. Here it is convenient to use the exponential representation of a complex scalar field,
\begin{equation}
 \phi=\rho{e^{i\alpha}}.
 \label{2}
\end{equation}
The Hamilton function of the original theory, that takes into account representation (\ref{2}), has the form
\begin{equation}
 H=\frac{1}{2}\int{d^3x}\left\{(E_i^2+B_i^2)+[p_\rho^2+\frac{p_\alpha^2}{\rho^2}+((\partial_i{\rho})^2+{\rho}^2(\partial_i{\alpha}-eA_i)^2)+V(\rho)]+A_0(ep_\alpha-\partial_iE_i)\right\}.
\end{equation}
The Gauss's law,
\begin{equation}
 ep_\alpha-\partial_iE_i=0,
 \label{4}
\end{equation}
serves as a constraint equation, and the scalar potential $A_0$ is a Lagrange multiplier. The gradient transformation affects the potentials of the electromagnetic field ($B_i$ remains unchanged) and the phase of the scalar field ($\rho$ remains unchanged). Infinitesimal transformation of the scalar potential
\begin{equation}
 \delta{A_0}=\dot{\epsilon}
 \label{5}
\end{equation}
compensates for the infinitesimal gradient transformation of the remaining physical variables, so that the action remains invariant. Therefore, we obtain the sought-for Euler-Lagrange equation for the resulting gauge variable in the physical sector of the theory by calculating the variation of the initial Lagrangian (\ref{1}) with infinitesimal variation (\ref{5}). This variation should be added to the original Lagrange function (1) as an additional condition. This is the introduction of a natural gauge condition into the original classical theory. Further, to simplify the consideration, we write explicitly only that part of the modified Lagrange function that is affected by the infinitesimal compensating transformation (\ref{5}):
\begin{equation}
 \tilde{L}=\frac{1}{2}\int{d^3x}\left\{[(\dot{A}_i-\partial_i{A_0})^2+{\rho}^2(\dot{\alpha}-eA_0)^2)]-(\dot{A}_i-\partial_i{A_0})\partial_i{\dot{\epsilon}}-e{\rho}^2(\dot{\alpha}-eA_0)\dot{\epsilon}\right\}.
 \end{equation}
 The remaining unchanged terms can be added at the very end of the consideration.

Passing to the canonical analysis of the modified theory, we write down the modified momenta,
\begin{equation}
  E_i=(\dot{A}_i-\partial_i{A_0})-\partial_i\dot{\epsilon}
 \end{equation}
 \begin{equation}
  P_\epsilon=\partial_i(\dot{A}_i-\partial_i{A_0})-e{\rho}^2(\dot{\alpha}-eA_0)
 \end{equation}
 \begin{equation}
  p_\alpha={\rho}^2(\dot{\alpha}-eA_0)-e{\rho}^2\dot{\epsilon}
 \end{equation}
 where $P_\epsilon$ is an additional momentum conjugate to the infinitesimal shift (\ref{5}). After that, we find the part of the Hamilton function of the modified theory that interests us:
 \begin{equation}
  \tilde{H}=\int{d^3x}[\tilde{h}+A_0(ep_\alpha-\partial_iE_i)],
 \end{equation}
 where
 \begin{equation}
  \tilde{h}=\frac{1}{2}[E_i^2+\frac{p_\alpha^2}{\rho^2}-(P_\epsilon+ep_\alpha-\partial_iE_i)(-\Delta+e^2\rho^2)^{-1}(P_\epsilon+ep_\alpha-\partial_iE_i)]
  \label{11}
 \end{equation}
 and $\Delta=\partial_i\partial_i$. We separated the constraint in the modified Hamilton function.
 Now the Hamilton function (\ref{11}) has a degenerate quadratic form of momenta, and the constraint (\ref{4}) plays the role of a condition on the initial data.

\section{Propagator of quantum electrodynamics in natural gauge}
Further, it is useful to analyze the consequences of the proposed modification in comparison with the usual approach to the construction of quantum theory. In the usual approach, based on the representation of the kernel of the evolution operator in the form of a functional integral on the phase space \cite{3}, we start by calculating the integral over momentum variables,

 \begin{equation}
  K=\int\prod{dA_0}\prod{d^3E}\prod{dp_\alpha}exp\left\{\frac{i}{\hbar}\int_0^t{dt}[\int{d^3x}(E_i\dot{A}_i+p_\alpha{\dot{\alpha}})-H]\right\}.
 \end{equation}
We omit constant factors as we integrate. There is no need to explicitly calculate this Gaussian integral over the momenta. The consequence of the additional integration over the Lagrange multiplier $A_0$ will be the degenerate nature of the kinetic energy - the quadratic function of velocities ($\dot{A}_i,\dot{\alpha }$). This is the reason for the divergence of the integral over the coordinates ($A_i,\alpha $) and an additional gauge condition is required to remove the degeneracy.
This will require an additional factor in the measure of integration over the coordinates ($A_i,\alpha $) (Faddeev-Popov determinant) so that the result does not depend on the choice of the gauge condition. As an example of such a condition, we take the Coulomb gauge condition:
\begin{equation}
  \partial_i{A_i}=0,
  \label{13}
 \end{equation}
 which, as will be seen, is the zero approximation of the natural gauge condition in the framework of perturbation theory. As is known, in quantum electrodynamics, this condition is not anomalous.

Let's move on to a modified theory. In this case, we write the kernel of the evolution operator in the form:
\begin{equation}
 \tilde{K}=\int\prod{dP_\epsilon{d\epsilon}}\prod{d^3E}\prod{dp_\alpha}exp\left\{\frac{i}{\hbar}\int_0^t{dt}[\int{d^3x}(E_i\dot{A}_i+p_\alpha{\dot{\alpha}}-\dot{P}_\epsilon{\epsilon})-\tilde{h}]\right\}.
 \label{14}
 \end{equation}
 The constraint is now not taken into account in evolution and there is no integration over the Lagrange multiplier $A_0$ in (\ref{14}). Instead, there is an integration over a pair of additional variables ($P_\epsilon,\epsilon$). But the result of this integration is trivial: The integral over $\epsilon$ gives $\dot{P}_\epsilon=0$, that is, the $P_\epsilon$ is integral of motion. Assuming it equal to zero, we will generally exclude additional variables from the modified theory.
  But they fulfilled their function - now the quadratic form of the momenta in Hamiltonian (\ref{11}) is degenerate. This means that the generalized velocities of the modified theory obey an additional condition, which is the desired natural gauge condition. To find it explicitly, we write down the conditions for the extremum of the modified action in the exponent in (\ref{14}) with respect to momenta ($E_i,p_\alpha$):
\begin{equation}
  \dot{A}_i-E_i-\partial_i(-\Delta+e^2{\rho}^2)^{-1}(ep_\alpha-\partial_k{E_k})=0,
 \end{equation}
 \begin{equation}
  \dot{\alpha}-\frac{p_\alpha}{\rho^2}-e(-\Delta+e^2{\rho}^2)^{-1}(ep_\alpha-\partial_k{E_k})=0.
 \end{equation}
 This extremum is found primarily when calculating the Gaussian integral over momenta \cite{3}. It is easy to verify that this implies:
 \begin{equation}
  \partial_i\dot{A}_i-e\rho^2\dot{\alpha}=0.
  \label{17}
 \end{equation}
 In the zero order of perturbation theory, this natural gauge condition preserves the Coulomb gauge (\ref{13}) in time.

To complete the integration procedure over the canonical momenta, we separate the transverse part of the electromagnetic field, and write its longitudinal part as:
\begin{equation}
  E_i^L=-\partial_i\lambda.
 \end{equation}
 We diagonalize the degenerate quadratic form of the momenta using the linear orthogonal transformations
  \begin{equation}
  s=ep_\alpha+\Delta\lambda,
 \end{equation}
 \begin{equation}
  r={A}p_\alpha+{B}\lambda.
 \end{equation}
where ${A}, {B}$ are still unknown Hermitian operators. In new variables ($s,r$), the quadratic form of the momenta takes the form:
\begin{eqnarray}
  \int{d^3}x[-(A^{-1}B+\Delta)^{-1}(A^{-1}r-s)\Delta(A^{-1}B+\Delta)^{-1}(A^{-1}r-s)
  \notag\\
  +\frac{1}{\rho^2}(A^{-1}+B^{-1}A)^{-1}(B^{-1}r+A^{-1}s)(A^{-1}+B^{-1}A)^{-1}(B^{-1}r+A^{-1}s)
  \notag\\
  -s(-\Delta+e^2\rho^2)^{-1}s].
  \label{21}
 \end{eqnarray}
 Degeneration means that there should be no dependence (\ref{21}) on the variable s. This reduces to two operator equations for finding the operators $A$ and $B$. We will not write them out here. We write down the resulting quadratic form of the momentum variable $r$ explicitly:
 \begin{eqnarray}
  (\hat{D}r,r)=\int{d^3x}[\frac{1}{\rho^2}(A^{-1}+B^{-1}A)^{-1}B^{-1}r(A^{-1}+B^{-1}A)^{-1}B^{-1}r
  \notag\\
  -(A^{-1}B+\Delta)^{-1}A^{-1}r\Delta(A^{-1}B+\Delta)^{-1}A^{-1}r].
  \label{22}
 \end{eqnarray}
The momentum $r$, together with the corresponding generalized coordinate, are obviously invariants of gradient transformations. When calculating the Gaussian integral with quadratic form (\ref{22}), the Faddeev-Popov determinant,
 \begin{equation}
  (det\hat{D})^{-1/2},
  \label{23}
 \end{equation}
appears, corresponding to the natural gauge condition (\ref{17}). We have obtained a convergent Feynman integral, which should be supplemented by the unchanged terms of the Hamilton function, omitted from the very beginning, and integration over the canonical coordinate conjugate to $r$ together with the corresponding Faddeev-Popov determinant (\ref{23}).

\section{Conclusions}
Thus, an alternative way of constructing the evolution operator of the gauge theory is proposed, in which the constraint equations are separated from the dynamics: the Hamilton function does not contain the gauge momenta corresponding to these constraints. Calculation of the Gaussian integral over momentum variables in the representation of the kernel of the evolution operator gives a natural gauge condition that restricts integration only over gauge-invariant canonical variables. Since the theory is modified at the classical level (there its dynamic content does not change), it is necessary to establish its correspondence with the usual approach after quantization.

\section{acknowledgements}
The authors thank V.A. Franke for useful discussions.

\end{document}